\documentclass[12pt]{article} 
\usepackage{graphics}
\usepackage{feynmp}
\usepackage{epsf}
\def\appendix#1{
  \addtocounter{section}{1}
  \setcounter{equation}{0}
  \renewcommand{\thesection}{\Alph{section}}
 \section*{Appendix \thesection\protect\indent \parbox[t]{11.715cm} {#1}}
  \addcontentsline{toc}{section}{Appendix \thesection\ \ \ #1}
  }

\newcommand{\newsection}{
\setcounter{equation}{0}
\section}

\catcode`\@=11 
 
\def\mycite{\@ifnextchar [{\@tempswatrue\@mycitex}{\@tempswafalse\@mycitex[]}} 
\def\mcite{\@ifnextchar [{\@tempswatrue\@mycitex}{\@tempswafalse\@mycitex[]}} 
 
\def\@mycitex[#1]#2{\if@filesw\immediate\write\@auxout{\string\citation{#2}}\fi 
 \def\@citea{}\@mycite{\@for\@citeb:=#2\do 
    {\@citea\def\@citea{,\penalty\@m\ }\@ifundefined 
       {b@\@citeb}{{\bf ?}\@warning 
       {Citation `\@citeb' on page \thepage \space undefined}}%
\hbox{\csname b@\@citeb\endcsname}}}{#1}}

\def\@mycite#1{[{#1}]} 
 
\catcode`\@=12 
\def\e{{\,\rm e}\,}
\newcommand{\rf}[1]{(\ref{#1})}
\newcommand{\non}{\nonumber \\*}
\newcommand{\be}{\begin{equation}}
\newcommand{\ee}{\end{equation}}
\def\bea{\begin{eqnarray}}
\def\eea{\end{eqnarray}}
\def\const{{\rm const}}
\def\la{\left\langle}
\def\ra{\right\rangle}
\def\d{\partial}
\newcommand\Tr{\mathop{\mathrm{Tr}}}
\newcommand\tr{\mathop{\mathrm{tr}}}
\newcommand\Sp{\mathop{\mathrm{Sp}}}

\def\D{\delta}
\def\dg{^\dagger}
\renewcommand{\/}{\!\!\!/}

\def\ep{\varepsilon}
\def\fpi{F_\pi}
\def\pp{\phi}
\def\tp{\theta}
\def\g{\Delta}
\def\gp{{\bf{\Delta}}}
\def\bpsi{\bar{\psi}}
\def\bchi{\bar{\chi}}
\def\gm{\gamma}
\def\S{\Sigma}
\def\seff{S_{\rm eff}}
\def\bp{{\bf{p}}}
\def\bn{{\bf{n}}}
\def\bk{{\bf{k}}}
\def\o{\omega}
\def\pone{{\bf{P_1}}{}}
\def\pe{{\bf{P_8}}{}}
\def\q{{\bf{Q}}}
\def\om{\Omega}
\def\a{\alpha}

\begin{document} 

\title{
\vspace{-0.5cm}
\hfill{\small ITEP-TH-9/00}\\
\vspace{0.1cm}
Dispersion Laws for Goldstone Bosons in a Color Superconductor \\ 
}
\author{\large  K. Zarembo\footnote{
Also at Institute for Theoretical and Experimental Physics,
 Moscow, Russia}
\vspace{0.1cm} \\ \small  
Department of Physics and Astronomy, 
\vspace{-0.2cm} \mbox{} \\ \small 
Pacific Institute for the Mathematical Sciences
\vspace{-0.2cm} \mbox{} \\ \small 
University of British Columbia
\vspace{-0.2cm} \mbox{} \\ \small 
Vancouver, BC V6T 1Z1,  Canada
\vspace{-0.1cm} \mbox{} \\ \small\tt
E-mail: zarembo@theory.physics.ubc.ca \\}
\date{}
\maketitle
\vspace{-1cm}
\begin{abstract}
\vspace{-0.2cm}
\baselineskip=12pt
The effective action for Goldstone bosons in the color-flavor locking
phase of dense QCD is analyzed. Interaction terms and higher
derivatives in the effective action appear to be controlled by
different scales. At energies of order of the superconducting gap,
 the derivative expansion
 breaks down,
while interactions still remain suppressed. The effective action 
valid at energies and momenta comparable to the gap is derived. 
Dispersion laws following from this action are such that
the energy of Goldstone bosons is always 
smaller than the gap in the quasiparticle spectrum, and
Goldstone bosons always propagate without damping.
\end{abstract}

\newpage\setcounter{page}{1}
\setcounter{equation}{0} 

\newsection{Introduction}

Cold and dense strongly interacting matter is expected to be in a
color superconducting state \cite{Bai84} 
at sufficiently high baryon density \cite{Alf97,Rap97}. 
The color superconducting phase is characterized by diquark condensate,
the actual structure of which depends on the number of quark
species whose masses are comparable to the gap in the
quasiparticle spectrum.
If three flavors can be considered light, which is definitely
true at very high density \cite{Alf99,Sch99}, 
the preferred ordering is color-flavor
locking (CFL) \cite{Alf98}:
\be
\la\psi^i_a\psi^j_b\ra\propto \ep^{ijk}\ep_{abk},
\ee
where $i,j$ and $a,b$ are the color and the flavor indices,
respectively.

The diquark condensate in the CFL phase breaks color $SU(3)_C$
and vector $SU(3)_V$ symmetries of QCD to the diagonal subgroup, and all
gluons except one acquire masses via the Higgs mechanism. Since one
  of the generators of $SU(3)_V$
is the electric charge, the photon mixes with one of the gluons to produce
a massless gauge field.
The baryon number $U(1)_B$, axial $U(1)_A$ and
chiral $SU(3)_A$ symmetries are broken by the diquark
condensate to ${\bf Z}_2$. 

Apart from the gauge boson of unbroken $U(1)$,
the only light degrees of freedom in the CFL phase 
at energies well below the superconducting gap
are Goldstone bosons of the broken global symmetries. 
The spectrum of
Goldstone bosons closely resembles the one in the hadronic phase
\cite{Alf98,Sch98,Pis99}, and,
following \cite{Son99}, 
the bosons associated with chiral symmetry below
will be referred to as pions and
the boson associated with $U(1)_A$ will be referred to as $\eta'$. 
Actually, $U(1)_A$ is not
exactly a symmetry of QCD because of the chiral anomaly, and $\eta'$ is not
exactly massless even in the chiral limit, 
but,  if the chemical potential is much larger than 
$\Lambda_{\rm QCD}$, the effects of anomalous violation
of $U(1)_A$ are negligibly small \cite{Sch99'}, and 
then $\eta'$ is sufficiently light
to be treated as a Goldstone boson.

The quark masses violate chiral symmetry and, consequently, 
make pions and $\eta'$ (but not to the baryon number
boson) massive. A usual way to deal with the quark masses is to treat them as
a perturbation\footnote{In contrast to the ordinary pion masses at zero
density, the squares of the Goldstone boson masses in the
CFL phase are 
proportional to the quark mass squared \cite{Alf98,Pis99}.}.
As a first approximation, quark masses can be neglected, and 
I will concentrate on the chiral limit throughout this paper. 

In the chiral limit, the effective action
for Goldstone bosons depends only on the derivatives of the
fields. Terms with
at most two derivatives are fixed by  
symmetries \cite{Hon99,Cas99}:
\bea\label{seff}
\mathcal{L}&=&
\frac{\fpi^2}{4}\tr\left( \d _0 U\dg\d _0 U-v_\pi^2\,\d_i U\dg\d_i U\right)
+\frac12\left( \d _0 \varphi\d _0 \varphi-v_B^2\,\d _i
 \varphi\d _i \varphi\right)\non &&
+\frac12\left( \d_0 \vartheta\d_0 \vartheta-v_{\eta'}^2\,\d_i 
\vartheta\d_i \vartheta\right),
\eea
where $$U=\exp\left(i\lambda^A\pi^A/\fpi\right),$$ $\lambda^A$ are
generators of $SU(3)$,  and the fields $\varphi$ and $\vartheta$
describe the baryon number and the $\eta'$ bosons. 

At large baryon density, the asymptotic
freedom allows to compute the parameters of the low-energy effective 
theory directly from QCD in a systematic way. 
In particular, all the parameters in \rf{seff}
 \cite{Son99,Rho99,Rho00}, as well as
corrections to the effective Lagrangian due to 
quark masses \cite{Son99,Hon99',Man00,Rho00,Bea00,Son00}, have been calculated.
The velocities of
all Goldstone bosons appear to be equal to the velocity of sound in the
relativistic fluid: $v^2=1/3$, and the pion
form-factor appears to be 
very large: $\fpi\sim\mu$. An important implication of
this fact is that pion-pion  interactions are weak
at low energies, because interaction vertices are suppressed
by powers of $\mu$, as can be seen expanding \rf{seff} in the pion
fields. 
There are two scales in the problem, however:
the chemical potential $\mu$ and the superconducting gap
$\g$, and $\g\ll\mu$. 
Higher-derivative corrections
to \rf{seff}, in general, 
are suppressed only by powers of the smaller scale $\g$, but
the interaction terms for all Goldstone bosons
appear to be suppressed by $\mu$ at any order of the derivative 
expansion, as will be shown 
in Sec.~\ref{gendis}. So, the Goldstone bosons remain weakly interacting at
energies comparable to the gap, while the kinetic terms in the
effective Lagrangian at these energies
are significantly changed by derivative corrections.
This means that the dynamics of the Goldstone bosons
remains relatively simple,  and
the effective theory for the Goldstone modes can be
extended to the energies comparable to the gap by taking into account
all orders of the derivative expansion. It should be mentioned that the 
effective theory becomes essentially non-local at energies and momenta
of order of the gap.

\newsection{General structure of the effective action}\label{gendis}

Let me sketch how the effective action for
the Goldstone bosons can be derived from the first principles.
At asymptotically large chemical potential, the dominant interaction
between quarks is the one-gluon exchange. The quark action
has the form:
\be\label{1}
S_{q}=\int d^4x\,\bpsi(i\d\/+\gm^0\mu)\psi-\frac{g^2}{2}
\int d^4x d^4y\,
 j^A_\mu(x)D^{\mu\nu}(x-y)j^A_\nu(y),
\ee
\be
j^A=\bpsi_a\gm_\mu T^A\psi_a.
\ee
Since condensation occurs in the diquark channel, it is natural to
represent the non-local four-fermion vertex in \rf{1} in the form
$$
\psi\dg_A\psi\dg_B G^{AB,CD}\psi_C\psi_D,
$$
where  color, flavor,
and  Dirac  
indices and space-time coordinates of the quark fields are
collectively denoted by one index. The four-fermion interaction
can be bosonized by Habbard-Stratonovich transformation, which
requires the introduction of the collective
field $\S^{ij}_{ab}(x,y)$ with the diquark quantum numbers.
The action
\be\label{hab}
S'_q=\int d^4x\,\bpsi(i\d\/+\gm^0\mu)\psi+\psi_A\S^{AB}\psi_B
+\psi\dg_A\S^{\dagger\,AB}\psi\dg_B
+\frac{2}{g^2}\,\S^{\dagger\,AB}G^{-1}_{AB,CD}\S^{CD}.
\ee
is then equivalent to \rf{1} after elimination of $\S$ via
its equations of motion. Alternatively,
integration over fermions yields the effective action for $\S$.
Schematically,
\be\label{effa}
\seff=-i\Tr\ln(i\d\/+\gm^0\mu+\S)+\frac{2}{g^2}\,\S\dg G^{-1}\S.
\ee
The vacuum expectation value of $\S$, which minimizes this action,
determines the superconducting gap.
The dominant gap is scalar, parity-even and anti-symmetric in color
\cite{Alf98}:
\be\label{vev}
\la\S^{ij}_{ab\,\alpha\beta}\ra=
(C\gm^5)_{\alpha\beta}\gp^{ij}_{ab},
\ee
\be
\gp^{ij}_{ab}=\g\ep^{ijk}\ep_{abk},
\ee
where $C=i\gm^2\gm^0$ is the charge conjugation matrix.
The value of the gap $\g$ is proportional to
$\e^{-\const/g}\mu$ \cite{Son98,Hon99'',Sch99'',Pis99'}.
The vacuum expectation value of $\S$ contains, in principle,
other Dirac, spin, and color-flavor structures, but all of them were
found to be suppressed at weak coupling 
\cite{Hsu99,Sch99',Sho99,Eva99,Bro99}.

The effective potential for $\S$ may have rather complicated structure, 
which is discussed
in \cite{Mir99,Sch99'}. But, irrespectively of the details of this
structure, the
minima of the potential are necessarily degenerate  due  to the
global symmetries of the original quark Lagrangian.
Any transformation of the form
\be\label{chir}
\S\rightarrow \e^{2i\pp+2i\gm^5\tp} V^T\S V,
\ee
where
$$
V=\e^{i\gm^5\lambda^A\pi^A/\fpi},
$$
leaves the action \rf{effa} invariant. The second term in the action 
is invariant because of the  $SU(3)_A\times U(1)_B\times U(1)_A$
symmetry of the gluon vertex. In the first term, the transformation
of the gap can be compensated by the rotation of the quark fields:
\be
\psi\rightarrow \e^{-i\pp-i\gm^5\tp}V\dg\psi.
\ee

Since the Goldstone modes correspond to motion along the degenerate minima
of the effective potential, the effective action for the Goldstone bosons 
is obtained by freezing
the modulus of $\S$ at its vacuum expectation value and allowing
the phases in \rf{chir} to depend on time and on space coordinates.
Since the gluon vertex is invariant even under local
$SU(3)_A\times U(1)_B\times U(1)_A$ transformations, the effective action
comes entirely from
the fermion determinant:
\be\label{s0}
S=-i\Tr\ln\left[i\d\/+\gm^0\mu+\e^{i\pp(x)+i\gm^5\tp(x)}V^T(x)\la\S(x,y)\ra
V(y)\e^{i\pp(y)+i\gm^5\tp(y)}\right].
\ee
The fields $\pp$ and $\tp$ here differ from $\varphi$ and $\vartheta$ 
in \rf{seff} by
normalization factors. 
Alternatively, $\S$ can always be aligned in a fixed direction by
rotation of the quark fields. Then the Goldstone fields appear in the
derivative term: 
\be\label{s1}
S=\Tr\ln\left(iD\/+\gm^0\mu+\la\S(x,y)\ra\right),
\ee
\be\label{covder}
D_\nu=\d_\nu+V\dg\d_\mu V+i\d_\nu\pp+i\gm^5\d_\nu\tp.
\ee
This form of the effective action was the starting point of 
Ref.~\cite{Son99}. 

An important feature of the effective action for the Goldstone bosons
is its independence of a particular form of the gluon vertex. If the
one-gluon exchange were replaced by any
$SU(3)_A\times U(1)_A\times U(1)_B$ invariant four-quark interaction,
the effective action would not change. 
In particular, the NJL model of \cite{Alf98} will lead to the same effective
theory for the Goldstone bosons as the one-gluon exchange
under the assumption of validity
of the mean field approximation, in other words, if the fluctuations
of $|\S|$ are neglected. This is also true for the instanton-based
models \cite{Alf97,Rap97,Car99,Rap99} with the exception that the quark
vertex  induced by instantons violates $U(1)_A$ and, hence, 
$\eta'$ should not be considered as a low-energy excitation.
Thus, the dynamics of the Goldstone bosons in the CFL phase is, to a
large extent, model independent, and the effective action \rf{s1} 
gives a reasonable description of the low-energy excitations
in the CFL phase independently of the underlying quark 
interactions, as long as fluctuations of $|\Sigma|$ are not too strong.

I would like to stress the analogy between
the effective theory for the Goldstone bosons in the color
superconductor \rf{s0} and the model of \cite{Dia88,Dia97}, 
in which the chiral Lagrangian for ordinary pions is  
induced by quark determinant. The pion fields  
enter through chiral rotations of the constituent quark mass. 
The constituent mass is the only dimensionful
parameter in the model \cite{Dia88,Dia97}
and sets the scale on which the derivative expansion of the
fermion determinant breaks down. In the case of the Goldstone bosons
in a color superconductor, the situation is different, as there are
two well separated scales:
 the gap $\g$ and the chemical potential
$\mu$, and $\mu\gg\g$. The presence of a small parameter $\g/\mu$ leads to
important simplifications. The derivative expansion is still
controlled by the mass gap, but, as will be shown shortly, the interaction
terms in the effective action are all suppressed by $1/\mu$.

\begin{center}
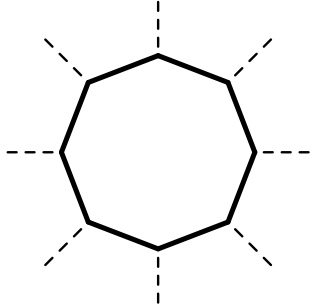
\begin{figure}[t]\begin{center}
\begin{fmffile}{fermionloop}
\begin{fmfgraph*}(40,40)
\fmfsurround{v1,v2,v3,v4,v5,v6,v7,v8}
\fmf{dashes}{v7,i7}
\fmf{dashes}{v2,i2}
\fmf{dashes}{v1,i1}\fmf{dashes}{v3,i3}\fmf{dashes}{v4,i4}\fmf{dashes}{v5,i5}
\fmf{dashes}{v6,i6}\fmf{dashes}{v8,i8}
\fmfcyclen{plain,width=thick}{i}{8}
\end{fmfgraph*}
\end{fmffile}\end{center}
\caption[x]{Typical diagram that contributes to the effective action.
\label{fig1}}
\end{figure}
\end{center}
The derivative expansion of \rf{s1} is given by a sum of quark loop 
diagrams of the  type shown in Fig.~\ref{fig1}. 
The diagram with $n$ external legs produces operators of dimension $n$
(and higher, if external momenta are taken into account). 
Coefficients before dimension $d$ operators are
dimension $(4-d)$ combinations of $\mu$ and $\g$. 
But the dependence
on $\mu$ can be easily found and actually is the same at any 
order of the 
derivative expansion. The reason is that, at large chemical potential,
 the momentum integral in the diagram \ref{fig1} 
is dominated by poles the quark propagators would have on the Fermi surface
in the absence of the gap. The leading order contribution to the
momentum integral comes from a thin shell around the Fermi surface:  
\be
\int\frac{d^4p}{(2\pi)^4}=\frac{\mu^2}{4\pi^3}\int dp_0 dq,
\ee
where $q=|\bp|-\mu$. If the integration over $p_0$ and $q$ converges,
the typical contribution comes from
$p_0,q\sim\g$ and does not depend on $\mu$.  
The factor of $4\pi\mu^2$ is merely the area of the Fermi
surface. 

The above arguments show that,
up to $O(\g/\mu)$ corrections, $\mu^2$ enters the effective
action only as an overall factor:
\be\label{form}
S=\mu^2\g^2\int d^4x\, \mathcal{L}
\left(\frac{U\dg\d U}{\g}\,,\,\frac{\d\pp}{\g}\,,\,\frac{\d\tp}{\g}\,;\,
\frac{\d^2}{\g^2}\right).
\ee
The conventional normalization of kinetic terms
in the action requires that the fields
are rescaled as: $U=\exp(i\lambda^A\pi^A/\mu)=1+i\lambda^A\pi^A/\mu
+\ldots$, $\pp=\varphi/\mu$ and $\tp=\vartheta/\mu$, 
after which all interaction terms appear to be suppressed by
powers of $1/\mu$. 
If all energies, momenta, and field strengths are much smaller than $\mu$,
 the effective action takes the form:
\be\label{qeff}
S=\int\frac{d^4k}{(2\pi)^4}\,\left[\pi^A(-k)\Pi\left(\frac{k^2}{\g^2}\right)
\pi^A(k)
+\varphi(-k) K\left(\frac{k^2}{\g^2}\right)\varphi(k)
+\vartheta(-k) \tilde{K}\left(\frac{k^2}{\g^2}\right)\vartheta(k)\right].
\ee
Further expansion in derivatives yields the 
low-energy effective Lagrangian \rf{seff},
but the effective action \rf{qeff} is valid also at
energies and momenta of order $\g$. In principle, interaction terms can be
computed perturbatively in $\g/\mu$.

Because of the lack of Lorentz invariance at finite
baryon density,  $k_0^2$ and $\bk^2$ enter the propagators of the
Goldstone bosons independently. The poles of the propagators
determine dispersion relations for the Goldstone bosons:
\be
\Pi(\o(\bk),\bk)=0,~~~~~K(\o_s(\bk),\bk)=0,~~~~~
\tilde{K}(\tilde{\o}_s(\bk),\bk)=0
\ee
The inverse
propagators $\Pi$, $K$, and $\tilde{K}$ are computed below in Sec.~\ref{3}.

\newsection{Interaction of quasiparticles with Goldstone \\ bosons}\label{2}

The gap in \rf{hab} mixes quarks with anti-quarks. The usual way to 
deal with that is to treat charge-conjugate 
quark operators,
\be
\psi_C=C\bpsi^T,
\ee
as independent fields.
The fermion part of the action \rf{hab} can be represented in the form
\be
S_f=\frac12\int d^4x\,\left[
\bpsi(i\d\/+\gm^0\mu)\psi+\bpsi_C(i\d\/-\gm^0\mu)\psi_C
+\bpsi_C\gm^5\gp\psi-\bpsi\gm^5\gp\psi_C
\right],
\ee
where $\S$ is fixed at its vacuum expectation value \rf{vev}.

Since the gap \rf{vev} does not mix left and right quasiparticles, it 
is convenient to
make one step more and to separate left and  right sectors
explicitly. Taking into account that
the charge conjugate of the right-handed spinor
is left-handed:
\be
(\psi_C)_L=\frac{1-\gm^5}{2}\,C\bpsi^T=(\psi_R)_C,
\ee
it is natural to  
describe left-handed and right-handed quasiparticles
by the following   four-com\-po\-nent spinors
\bea
\chi_L&=&\frac{1-\gm^5}{2}\,\psi+\frac{1+\gm^5}{2}\,\psi_C\,,
\non
\chi_R&=&\frac{1+\gm^5}{2}\,\psi+\frac{1-\gm^5}{2}\,\psi_C\,.
\eea
The Lagrangian decouples into two independent pieces 
in terms of $\chi_{L}$ and $\chi_{R}$:
\be\label{lagfer}
\mathcal{L}_f=\frac12\,\bchi_L(i\d\/-\gm^0\gm^5\mu-\gp)\chi_L
+\frac12\,\bchi_R(i\d\/+\gm^0\gm^5\mu+\gp)\chi_R.
\ee
The fields $\chi_L$ and $\chi_R$ are then treated as independent variables
in the path integral.

The interaction of quasiparticles with gluons and Goldstone bosons is
described by the Lagrangian
\be
\mathcal{L}_f=\frac12\,\bchi_L(iD\/^L-\gm^0\gm^5\mu-\gp)\chi_L
+\frac12\,\bchi_R(iD\/^R+\gm^0\gm^5\mu+\gp)\chi_R.
\ee
The form of the covariant derivatives $D^L$ and $D^R$ is dictated by
transformation laws of $\chi_L$ and $\chi_R$
under chiral, gauge, and baryon number transformations.
As follows from the definition of $\chi_L$, $\chi_R$
in terms of the original quark operators, these fields transform as
\bea
\chi_{R}&\rightarrow&\e^{-i\gm^5(\tp+\pp)}
\left(\frac{1+\gm^5}{2}\,U\dg\om\dg
+\frac{1-\gm^5}{2}\,U^T\om^T\right)\chi_R,
\non
\chi_{L}&\rightarrow&\e^{-i\gm^5(\tp-\pp)}
\left(\frac{1+\gm^5}{2}\,U^*\om^T
+\frac{1-\gm^5}{2}\,U\om\dg\right)\chi_L,
\eea
where $\om$ acts on color indices and $U$ describes
the rotation in the flavor space.
The above transformation laws yield for
covariant derivatives:
\bea\label{dr}
D^R_\mu&=&\d_\mu+\frac12(L_\mu-L_\mu^T)+\frac{g}{2}(A_\mu-A_\mu^T)
\non &&
+\gm^5\left[\frac12(L_\mu+L_\mu^T)+\frac{g}{2}(A_\mu+A_\mu^T)
+i\d_\mu\tp+i\d_\mu\pp\right],
\\*\label{dl}
D^L_\mu&=&\d_\mu-\frac12(R_\mu-R_\mu^T)+\frac{g}{2}(A_\mu-A_\mu^T)
\non &&
+\gm^5\left[\frac12(R_\mu+R_\mu^T)-\frac{g}{2}(A_\mu+A_\mu^T)
+i\d_\mu\tp-i\d_\mu\pp\right],
\eea
where $g$ is the QCD coupling, $A_\mu$ is the anti-Hermitian gluon field, 
$L_\mu$ and $R_\mu$ are the left and the right pion currents:
\be
L_\mu=U\dg\d_\mu U=\frac{i}{\fpi}\,\d_\mu\pi+O(1/\mu^2),
~~~~~R_\mu=\d_\mu UU\dg=\frac{i}{\fpi}\,\d_\mu\pi+O(1/\mu^2).
\ee

The quasiparticle propagators can be found 
by diagonalization of the Dirac operators in \rf{lagfer}.
The color-flavor structure of the gap matrix 
can be diagonalized by separating components of $\chi$ belonging to
the octet  and the singlet representations of the unbroken $SU(3)$
group. The projectors on definite representations have the form:
\be
\pone^{\,ij}_{\,ab}=\frac13\,\D^i_a\D^j_b,
\ee
\be
\pe^{\,ij}_{\,ab}=\D^{ij}\D_{ab}-\frac13\,\D^i_a\D^j_b.
\ee 
The gap matrix can be decomposed as
\be
\gp=\g\q+2\g\pone,
\ee
where 
\be
\q^{ij}_{ab}=\frac13\,\D^i_a\D^j_b-\D^i_b\D^j_a
\ee
is the square root of $\pe$:
\be
\q^{ij}_{ab}\q^{jk}_{bc}=\pe^{\,ik}_{\,ac}.
\ee
The Dirac structure of the
quasiparticle Lagrangian is diagonalized with the help of
the helicity projector:
\be
H_\pm=\frac{1\pm \gm^0\gm^5\gm^in_i}{2}\,,~~~~~\bn=\frac{\bp}{|\bp|}.
\ee
The propagator of right quasiparticles is
\be\label{prop}
S_R(p)=H_+\left(\frac{p\/_--\g\q}{p_-^2-\g^2}\,\pe
+\frac{p\/_--2\g}{p_-^2-4\g^2}\,\pone\right)
+H_-\left(\frac{p\/_+-\g\q}{p_+^2-\g^2}\,\pe
+\frac{p\/_+-2\g}{p_+^2-4\g^2}\,\pone\right),
\ee
where
\be
p_\pm=(p_0,\bp\pm\mu\bn)=(p_0,(|\bp|\pm\mu)\bn).
\ee
The propagator for the left quasiparticles is obtained by
flipping signs of $\mu$ and $\g$. 

Only the first term in the propagator picks up large 
contributions at the Fermi surface, and the second term will be
omitted in most of the calculations below. In principle, a separation
of the relevant modes  could have been done
directly on the level of the action for quasiparticles \cite{Hon98}.

\newsection{Effective action and dispersion of the Goldstone
bosons}\label{3}

As discussed in Sec.~\rf{gendis}, interactions of the Goldstone bosons are
suppressed at large chemical potential, so the quadratic terms constitute
the most important part of the effective action
\be
S=-\frac{i}{2}\,\Tr\ln(iD\/^L-\gm^0\gm^5\mu-\gp)
-\frac{i}{2}\,\Tr\ln(iD\/^R+\gm^0\gm^5\mu+\gp).
\ee
The calculation of the quadratic terms in the effective action
amounts to evaluation of polarization
operators which
arise from the derivative expansion of fermion determinants:
\be\label{det}
-\frac{i}{2}\,\Tr\ln(iD\/\pm\gm^0\gm^5\mu\pm\gp)
=-\tr v_\mu\Pi^{\mu\nu}v_\nu-\tr a_\mu\Pi^{\mu\nu}a_\nu+
s_\mu K^{\mu\nu}s_\nu+\ldots.
\ee
Here, the covariant derivative
\be
D_\mu=\d_\mu+v_\mu+\gm^5(a_\mu+is_\mu)
\ee
contains $SU(3)$ singlet $s_\mu$ and adjoint fields
$a_\mu$, $v_\mu$, that
 obey 
\be\label{symm}
a_\mu^T=a_\mu,~~~~~v_\mu^T=-v_\mu.
\ee
The polarization
operators are even functions of $\mu$ and $\g$ and, thus, are the same for
the left and the right quasiparticles. It is less obvious that the axial
and the vector polarization operators are the same, but it also follows
from the symmetry properties \rf{symm} and the color-flavor structure of the 
quasiparticle propagator. 

Substitution of the explicit expressions for the covariant 
derivatives \rf{dr}, \rf{dl}  gives for 
the terms quadratic in currents:
\be
S=-\tr L_\mu\Pi^{\mu\nu}L_\nu-\tr R_\mu\Pi^{\mu\nu}R_\nu
+\d_\mu\tp \,K^{\mu\nu}\,\d_\nu\tp 
+\d_\mu\pp \,K^{\mu\nu}\,\d\pp_\nu+\ldots\,.
\ee 
Consequently,
the inverse propagators of the Goldstone bosons are given by
longitudinal components of the polarization operators: 
\be\label{prodol}
\Pi(k)=k_\mu\Pi^{\mu\nu}(k)k_\nu,~~~~~
\tilde{K}(k)=K(k)=k_\mu K^{\mu\nu}(k)k_\nu.
\ee

The same polarization operators describe mass terms for gluons and
the mixing between gluons and the Goldstone boson currents. The
appearance of mixed terms does not imply mixing of
gluons with pions, which is forbidden by parity.
$A_\mu$  mixes with the difference of currents
 $L_\mu-R_\mu$ that is at most quadratic in pion fields, and 
the mixing describes parity-even
interaction:
\be
\mathcal{L}_{g\pi\pi}\propto g\mu^2 \tr A^\nu(L_\mu-R_\mu).
\ee
This term arises because pions are colored 
as a result of the color-flavor locking. It can be rewritten as
\be
\mathcal{L}_{g\pi\pi}\propto g\mu^2 \tr\d_\nu U[A^\nu,U\dg],
\ee
which is a part of the covariant derivative squared,
$$
\tr (D_\mu U)\dg D^\mu U,~~~~~D_\mu=\d_\mu+g[A_\mu,\cdot].
$$
The appearance of this term  means that 
ordinary derivatives in the effective Lagrangian \rf{seff} 
must be replaced by covariant ones,
if the gluon fields are taken into account.

The polarization operators are calculated explicitly in the Appendix.
Extracting the longitudinal part from
Eqs.~\rf{pi1}--\rf{pi3} we get, according to \rf{prodol}, 
the inverse pion propagator:
\bea\label{pionpr}
\Pi(\o,k)&=&
-\frac{\mu^2\g^2}{24\pi^2k}\int_0^1dx\,\int_0^kd\a\,(\a^2-\o^2)\left[
\frac{5}{x(1-x)(\a^2-\o^2)+\g^2}
\right.\non && \left.
+\frac{4-6x}{x(1-x)(\a^2-\o^2)+(4-3x)\g^2}
\right],
\eea
where $\o=k_0$ and $k=|\bk|$.
Expansion at low momenta reproduces the result of
\cite{Son99} for the pion form-factor:
\bea
\Pi(\o,k)&=&
\frac{\mu^2}{24\pi^2}\left[\frac13\,\left(21-8\ln 2\right)
\left(\o^2-\frac13\,k^2\right)
\right.\non &&\left.
+\frac{1}{54\g^2}(123-112\ln 2)
\left(\o^4-\frac23\,k^2\o^2+\frac15\,k^4\right)
+O\left(\frac{1}{\g^4}\right)\right].
\eea

\begin{center}
\begin{figure}[t]
\epsfysize=6cm
       \epsffile{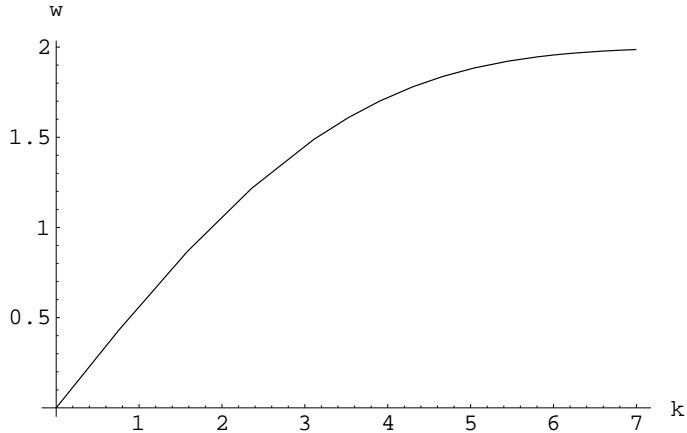}
       \caption[x]{$\protect\omega$ as a function of $k$ in the units of 
$\protect\Delta$.\label{disp}}
       \end{figure}
\end{center} 

The pole of the propagator determines the dispersion law for 
pions.
At low momenta, pions have the same dispersion relation as relativistic
density waves \cite{Son99}. The corrections tend to decrease the frequency:
\be\label{dispk0}
\o(k)=\frac{1}{\sqrt{3}}\,k\left[1-\frac{1}{135}\,
\frac{123-112\ln 2}{21-8\ln 2}\,\frac{k^2}{\g^2}
+O\left(\frac{k^4}{\g^4}\right)\right]~~~~~(k\rightarrow 0).
\ee
At higher momenta, the dispersion curve becomes more steep and at 
$k\rightarrow\infty$ the frequency approaches $2\g$ from below,
since:
\be\label{largek}
\Pi(\o,k)\approx
\frac{\mu^2\g^2}{24\pi^2}\left[\frac{10\pi\g}{k}\,
\ln\left(\frac{\g}{2\g-\o}\right)
-24\ln \left(\frac{k}{\g}\right)\right]
\ee
at $\o\rightarrow 2\g$ and $k\rightarrow\infty$. Consequently,
\be
\o(k)=2\g-\const\,\,\g
\exp\left[-\frac{12k}{5\pi\g}\,\ln \left(\frac{k}{\g}\right)
\right]~~~~~(k\rightarrow\infty).
\ee
Here,
the momentum $k$ is assumed to be large compared to $\g$, but
should be much smaller than $\mu$. Actually, this result can be
trusted up to $k\sim\sqrt{\g\mu}$, after which the approximation
\rf{expan} used in its derivation is no longer valid\footnote{I am
grateful to D.K.~Hong and K.~Rajagopal for the discussion of 
this point.}.

The dispersion curve for pions is shown in Fig.~\ref{disp}. 
The energy of a pion never becomes larger than the gap in the quasiparticle 
spectrum, and the decay of the pion to the constituent 
quasiparticle-anti-quasiparticle pair is always kinematically forbidden.
The pion frequency has no imaginary part, as a result, and 
pions propagate without damping, unless weak interactions are taken
into account.

The energy of a pion can never reach $2\g$, because
the polarization operator $\Pi^{\mu\nu}(\o,\bk)$ has a logarithmic
singularity when the energy approaches the threshold
of the pair creation. The singularity comes from the
integration over small $\a=\bk\cdot\bp/|\bp|$, where $\bp$ is the momentum
of quasiparticles in the loop. Thus, the threshold singularity is due to 
creation of an almost on-shell Cooper pair with the  momentum of quasiparticles
perpendicular to the momentum of the pion. This singularity compensates for the
second term in \rf{largek}, which comes from integration over $\a\sim k$, and
thus corresponds to creation of the Cooper pair composed of quasiparticles
moving in the same direction as the pion.
The appearance of the logarithmic
singularity is a manifestation of the (2+1)-dimensional nature
of the momentum integration near the Fermi surface. In (3+1) dimensions,
the square root singularity would arise, which does not blow up at the
threshold.

The fact that pions can mix with virtual Cooper pairs might seem puzzling, since a 
Cooper pair has baryon charge 2. There is no contradiction with baryon charge
conservation, however, as we are dealing with the unusual situation in
which the baryon symmetry is broken. The vacuum, as well as excited states,
are not eigenstates of the baryon charge operator, so the pions do not have
definite baryon charge and symmetry arguments do not forbid a non-zero matrix element
$\left\langle {\rm vac}| \psi\psi |\pi\right\rangle$.

The inverse propagator of the singlet Goldstone bosons follows from
Eqs.~\rf{k1}--\rf{k3}:
\bea
K(\o,k)&=&-\frac{\mu^2\g^2}{\pi^2k}\int_0^1dx\,\int_0^kd\a\,
(\a^2-\o^2)\left[
\frac{2}{x(1-x)(\a^2-\o^2)+\g^2}
\right.\non && \left.
+\frac{1}{x(1-x)(\a^2-\o^2)+4\g^2}
\right].
\eea
The dispersion relation is qualitatively the same as that for the octet
Goldstone bosons. At low momenta:
\be
\o_s(k)=\frac{1}{\sqrt{3}}\,k\left[1-\frac{11}{540}\,
\frac{k^2}{\g^2}
+O\left(\frac{k^4}{\g^4}\right)\right]~~~~~(k\rightarrow 0).
\ee
At high momenta:
\be
\o_s(k)=2\g-\const\,\,\g
\exp\left[-\frac{3k}{\pi\g}\,\ln \left(\frac{k}{\g}\right)
\right]~~~~~(k\rightarrow\infty).
\ee

\newsection{Discussion}\label{4}

The order parameter of the color superconductivity is the diquark
condensate or the conjugate variable, the gap. The orientation of
the condensate in the  color and flavor space describe the Goldstone
modes, which are the only low-energy
degrees of freedom away from the phase transition.
Power-counting arguments show that the Goldstone bosons are
weakly interacting due to the presence of two widely separated scales in
the problem: $\g\ll\mu$. The effective action for Goldstone bosons has
a form \rf{form} and, after appropriate rescaling of the fields,
all nonlinearities in it can be omitted. 
This approximation, of course, implies that the Goldstone
fields themselves are not too strong: $|\pi^A|\sim\g$.
The situation is, to some extent, inverse to what happens 
at zero chemical potential. Interaction terms and derivative
corrections
in the ordinary chiral Lagrangian become important more or less
at the same scale. The parameter that governs interactions ($\fpi$)
may even be smaller (cf.~\cite{Dia88,Dia97}), so the chiral dynamics
is essentially non-linear. In the CFL phase, the non-linear
effects are much less important than derivative corrections.

The simplicity of the dynamics of the Goldstone bosons allows to go beyond
the derivative expansion and to derive the 
effective action for the Goldstone bosons which contains
all powers of derivatives and is valid at 
energies and momenta of order of the gap. 
This effective action determines the dispersion laws for the Goldstone
bosons. The dispersion of the Goldstone bosons
turns out to be similar to the
dispersion of phonons in a crystal. In particular, 
the energy of Goldstone bosons does not grow indefinitely as  a
function of momentum and is always smaller than the gap in the
quasiparticle spectrum. The dispersion curve flattens at large momenta and 
the group velocity of Goldstone bosons becomes exponentially small.

The deviation of the velocity of pions from the velocity of light by
a factor of $1/\sqrt{3}$ has interesting implications for weak
interactions in a color superconductor. Pions in a CFL phase are
much lighter than in the vacuum: $m_\pi^2\propto m_sm_{u,d}\g^2/\mu^2$
[possibly, up to a factor of $\ln(\mu/\g)$]  
 \cite{Hon99',Man00,Rho00,Bea00,Son00}. 
With  $\g/\mu\approx 1/10$ taken as an order of magnitude estimate,
$m_e\ll m_\pi\ll m_\mu$, so the dominant decay mode of charged pions is
\be\label{pienu}
\pi^\pm\rightarrow e^\pm\nu,
\ee
For this decay to be kinematically allowed, 
the energy of a pion must at least be larger than the energy of an
electron with the same momentum: $\omega(k)>\sqrt{k^2+m_e^2}$.
Since the energy of a pion is bounded by $2\g$, the decay is forbidden
at large enough $k$. 
In fact, the critical momentum
is much smaller than $\g$: With the mass of the pion taken into
account, the dispersion relation \rf{dispk0} becomes
\be
\omega(k)=\sqrt{\frac13\,k^2+m_\pi^2}.
\ee
The energy conservation then requires
$$
\frac13\,k^2+m_\pi^2>k^2+m_e^2,
$$
or, neglecting electron mass,
\be
k<k_c=\sqrt{\frac{3}{2}}m_\pi.
\ee
Charged pions moving with the momentum $k>k_c$ will be absolutely stable.
The decay \rf{pienu} then will go in the opposite direction:
\be
e^\pm\rightarrow\pi^\pm\nu.
\ee
Therefore, fast electrons will be converted into pions in a color
superconductor.

For similar reasons, an on-shell photon can always decay into
baryon number Goldstone bosons, which are exactly massless.

\subsection*{Acknowledgements}
I am grateful to D.K.~Hong, K.~Rajagopal,
T.~Sch\"afer and A.~Zhitnitsky for discussions.
The work was supported by  
PIms Postdoctoral Fellowship and by NSERC of Canada.

\setcounter{section}{0}

\appendix{Polarization operators}

There are two contributions of order $\mu^2$ 
to the polarization operators defined in \rf{det}. One comes from
momenta close to the Fermi surface:
$q\equiv|\bp|-\mu\ll\mu$. 
Using quasiparticle propagators \rf{prop} and symmetry
properties \rf{symm}, we find for
the octet polarization operator in \rf{det}:
\bea\label{loop}
\Pi^{\mu\nu}&=&\frac{i}{12}\int\frac{d^4p}{(2\pi)^4}\,
\left\{\frac{7J^{\mu\nu}+2\g^2I^{\mu\nu}}{[(p+k)_-^2-\g^2](p_-^2-\g^2)}
\right.\non && \left.
+\frac{2J^{\mu\nu}+4\g^2I^{\mu\nu}}{[(p+k)_-^2-\g^2](p_-^2-4\g^2)}
\right\}
+\D\Pi^{\mu\nu},
\eea
where
\be
I^{\mu\nu}=\Sp\gm^\mu H_+\gm^\nu H_+,
\ee
\be
J^{\mu\nu}=\Sp\gm^\mu H_+(p\/+k\/)_-\gm^\nu H_+p\/_-.
\ee
Only positive-helicity parts of the quasiparticle propagartors, which are
singular at the Fermi surface, were kept in \rf{loop},
since typical $p_0$ and
$q$ that contribute to the loop integral are of order of $\g$.

The term with two opposite helicity
projectors, though not singular at the Fermi surface, is quadratically
divergent and 
receives $O(\mu^2)$ contribution from the modes with $q\sim\mu$,
which means that
the dependence on $\g$ and $k\sim\g$ can be omitted from this term:
\be
\D\Pi^{\mu\nu}=\frac{3i}{2}\int\frac{d^4p}{(2\pi)^4}\,\left(
\frac{\Sp\gm^\mu H_+p\/_-\gm^\nu H_-p\/_+}{p_-^2p_+^2}
-\frac{\Sp\gm^\mu p\/\gm^\nu p\/}{p^4}\right).
\ee
Subtraction of the polarization operator at zero chemical potential
serves as a gauge-invariant regularization. A simple algebra yields:
\be
\D\Pi^{00}=0,
\ee
\be
\D\Pi^{ij}=-\D_{ij}\int\frac{d^3p}{(2\pi)^3}\,\frac{1}{|\bp|}
\theta(|\bp|<\mu)
=-\frac{\mu^2}{4\pi^2}\,\D^{ij}.
\ee

Below, $k_0$ is denoted by $\o$ and $|\bk|$ is denoted by $k$. 
Both are much smaller than $\mu$, and approximations like
\be\label{expan}
|\bp+\bk|-\mu\approx q+\bn\cdot\bk
\ee
greatly simplify the Dirac traces and the momentum integral in \rf{loop}.
Omitting $O(\g/\mu)$ corrections:
\be
I^{00}=2,~~~~~I^{ij}=-2n^in^j,
\ee
\be
J^{00}=2\left[p_0(p_0+\o)+q(q+\a)\right],
\ee
\be
J^{0i}=4(p_0+\o)qn^i,~~~~~J^{i0}=4p_0(q+\a)n^i,
\ee
\be
J^{ij}=2n^in^j\left[p_0(p_0+\o)+q(q+\a)\right].
\ee
Here $\a=\bn\cdot\bk$.

The momentum integral in \rf{loop} can be done 
by Wick rotation to the Euclidean space and subsequent use of the
Feynman parameterization. The energy integral should be done first
\cite{Son99},
which yields:
\bea\label{pi1}
\Pi^{00}&=&\frac{\mu^2}{24\pi^2k}\int_0^1dx\,\int_0^kd\a\,\left[
9+\frac{7x(1-x)(\a^2+\o^2)-2\g^2}{x(1-x)(\a^2-\o^2)+\g^2}
\right.\non && \left.
+\frac{2x(1-x)(\a^2+\o^2)-4\g^2}{x(1-x)(\a^2-\o^2)+(4-3x)\g^2}
\right],
\eea
\bea
\Pi^{0i}=\Pi^{i0}&=&\frac{\mu^2\o k^i}{12\pi^2k^3}\int_0^1dx\,\int_0^kd\a\,
\left[
\frac{7x(1-x)\a^2}{x(1-x)(\a^2-\o^2)+\g^2}
\right.\non && \left.
+\frac{2x(1-x)\a^2}{x(1-x)(\a^2-\o^2)+(4-3x)\g^2}
\right],
\eea
\bea\label{pi3}
\Pi^{ij}&=&\frac{\mu^2}{24\pi^2k}\int_0^1dx\,\int_0^kd\a\,\left\{
3\D^{ij}+\frac{1}{2}\left[\left(1-\frac{\a^2}{k^2}\right)\D^{ij}
+\left(3\frac{\a^2}{k^2}-1\right)\frac{k^ik^j}{k^2}\right]
\right.\non && \left.
\times\left[
\frac{7x(1-x)(\a^2+\o^2)+2\g^2}{x(1-x)(\a^2-\o^2)+\g^2}
+\frac{2x(1-x)(\a^2+\o^2)+4\g^2}{x(1-x)(\a^2-\o^2)+(4-3x)\g^2}\right]
\right\}.
\eea
It is straightforward to check that the polarization operator
is transverse.

The color-flavor algebra is simpler for singlet polarization operator,
since in that case the quasiparticle in the loop belongs to 
a definite representation of
$SU(3)$. The trace over color and flavor indices gives just the
dimension of the representation:
\bea\label{loop1}
K^{\mu\nu}&=&\frac{i}{4}\int\frac{d^4p}{(2\pi)^4}\,
\left\{8\,\frac{J^{\mu\nu}-\g^2I^{\mu\nu}}{[(p+k)_-^2-\g^2](p_-^2-\g^2)}
\right.\non && \left.
+\frac{J^{\mu\nu}-4\g^2I^{\mu\nu}}{[(p+k)_-^2-4\g^2](p_-^2-4\g^2)}
\right\}
+3\D\Pi^{\mu\nu}.
\eea
The coefficient $3$ before the high-momentum contribution is due to the
extra trace over flavor indices.

The momentum integration yields:
\bea\label{k1}
K^{00}&=&\frac{\mu^2}{8\pi^2k}\int_0^1dx\,\int_0^kd\a\,\left[
9+8\,\frac{x(1-x)(\a^2+\o^2)+\g^2}{x(1-x)(\a^2-\o^2)+\g^2}
\right.\non && \left.
+\frac{x(1-x)(\a^2+\o^2)+4\g^2}{x(1-x)(\a^2-\o^2)+4\g^2}
\right],
\eea
\bea
K^{0i}=K^{i0}&=&\frac{\mu^2\o k^i}{4\pi^2k^3}\int_0^1dx\,\int_0^kd\a\,
\left[8\,
\frac{x(1-x)\a^2}{x(1-x)(\a^2-\o^2)+\g^2}
\right.\non && \left.
+\frac{x(1-x)\a^2}{x(1-x)(\a^2-\o^2)+4\g^2}
\right],
\eea
\bea\label{k3}
K^{ij}&=&\frac{\mu^2}{8\pi^2k}\int_0^1dx\,\int_0^kd\a\,\left\{
3\D^{ij}+\frac{1}{2}\left[\left(1-\frac{\a^2}{k^2}\right)\D^{ij}
+\left(3\frac{\a^2}{k^2}-1\right)\frac{k^ik^j}{k^2}\right]
\right.\non && \left.
\times\left[
8\,\frac{x(1-x)(\a^2+\o^2)-\g^2}{x(1-x)(\a^2-\o^2)+\g^2}
+\frac{x(1-x)(\a^2+\o^2)-4\g^2}{x(1-x)(\a^2-\o^2)+4\g^2}
\right]\right\}.
\eea
This polarization operator is also transverse, as it should be.

\end{document}